\documentclass[a4paper]{jpconf}
\usepackage[english]{babel}

\usepackage{rotating}

\usepackage{caption}

\usepackage{graphicx}

\usepackage{color}
\newcommand{\pmn}{PbMg$_{1/3}$Nb$_{2/3}$O$_{3}$\,}
\begin{document}
\title{Study of diffuse scattering under hydrostatic pressure in PbMg$_{1/3}$Nb$_{2/3}$O$_{3}$}

\author{G-M Rotaru$^1$, B Padmanabhan$^1$, S N Gvasaliya$^1$, B Roessli$^1$, Th Str\"{a}ssle$^1$, R A Cowley$^2$, S G Lushnikov$^3$,
S. Klotz$^4$}
\address{$^1$ Laboratory for Neutron Scattering, ETH Zurich and Paul Scherrer Institut, CH-5232 Villigen, Switzerland}
\address{$^2$ Clarendon Laboratory, Oxford University, Parks Road, Oxford OX1 3PU, UK}
\address{$^3$ Ioffe Physical Technical Institute, 26 Politekhnicheskaya, 194021 St Petersburg, Russia}
\address{$^4$ Physique des Milieux Denses, IMPMC, CNRS UMR 7590, Universit\'{e} P. M. Curie, 4 Place Jussieu, 75252 Paris, France}
\ead{gelu.rotaru@psi.ch}

\begin{abstract}
We report measurements of the evolution of the diffuse scattering in  a single crystal \pmn  as a function of hydrostatic 
pressure. Upon applying pressure the diffuse scattering intensity decreases and is suppressed at about 3~GPa, while no change in the line shape is observed.
Correlations between Pb displacements, diffuse scattering and relaxor properties are discussed.

\end{abstract}

\section{Introduction}

Relaxors are of scientific and commercial interest  due to their dielectric properties.
The giant piezoelectric and electrostriction effects present in these materials are exploited, for example, in sensors and actuators.
\pmn (PMN) is a well-known complex perovskite that exhibits relaxor ferroelectric properties. 
The average structure stays centrosymmetric with the space group
Pm3m down to low temperatures although the dielectric permittivity has a broad maximum around 270~K when measured at 1~kHz~\cite{nawrocik}. 
It has been shown that the lowest transverse optic mode does not condense at this temperature~\cite{stock},~\cite{hlinka} as  happens in the classical 
ferroelectrics at the structural phase transition. On the other hand a quasielastic mode was found whose susceptibility 
has a similar behavior with the temperature dependence of the dielectric permittivity \cite{jpcm2005}.
 The current interpretation is that the dielectric properties in PMN   
find their origins in nanoscale correlated regions, the so-called polar nano-regions (PNR). These polar nanoregions appear 
at  the Burns temperature of about  620~K   and grow when the temperature is lowered. Evidence for the formation of PNR in 
relaxor ferroelectrics comes from the presence of temperature dependent anisotropic diffuse scattering close to the Bragg
 reflections in X-rays and neutron scattering experiments. It is generally accepted that these are correlated displacements
 of Pb-ions that are mainly responsible for the formation of PNR. Systematic studies of diffuse scattering as a function
 of temperature have been reported, e.g., in Refs~\cite{jpcm2005},~\cite{matsuura},~\cite{xu}. 

Another parameter that changes the dielectric properties is the hydrostatic pressure~\cite{samara}.
Pressure reduces the peak in the dielectric permittivity and shifts its maximum towards lower temperatures \cite{nawrocik}.
Suppression of the intensity of the X-ray diffuse scattering~\cite{chaabane} and  structural phase transitions
 are reported in PMN at about 4.5~GPa~\cite{chaabane},~\cite{raman},~\cite{brillouin} and in PbSc$_{1/2}$Ta$_{1/2}$O$_{3}$ at 1.9~GPa ~\cite{pst}.
Neutron powder diffraction experiments \cite{pmnpressure} show that the atomic displacements of Pb diminish 
and anisotropy of oxygen temperature factor increases with pressure. 
The aim of the present paper is to study the neutron diffuse scattering as a function of pressure in a single crystal of PMN.

\section{Experimental details}
The measurements were carried out with the cold neutron three-axis spectrometer TASP~\cite{tasp} at SINQ. 
For this experiment a $~$3x3x2~mm$^{3}$ single crystal of PMN was oriented with the [1,0,0] and [0,1,0] crystallographic 
directions in the scattering plane. The use of the spectrometer was justified by the need to reduce the background and hence
to improve the signal-to-noise ratio. The spectrometer was configured in elastic mode with k$_f = 1.97$~\AA $^{-1}$, collimation open-$-80'-80'-80'$.

To apply pressure, the crystal was embedded in a lead matrix placed into a Cu-Be gasket. Here, lead acts as pressure transmitting medium.
Then the crystal~-~gasket assembly  was loaded in a Paris-Edinburgh cell \cite{paris_edinburgh}. The entire pressure cell is mounted in
a cryostat. The lead matrix in which the sample is embedded is also used as a pressure calibrant, the pressure being determined
from the change in volume of the Pb unit cell using the third-order Birch-Murnaghan equation of state \cite{birch}. 
As the diffuse scattering in PMN is most intense at lower temperatures \cite{jpcm2005} the sample was cooled down to 80K. 
 The spectra were collected in the (1,0,0) Brillouin zone.
 
\section{Results}
We have measured the diffuse scattering distribution around the (1,0,0) Bragg reflection at 0.5~GPa and 2.2~GPa. 
Systematic measurements of the diffuse scattering as a function of pressure were performed along the $[1,-1,0]$ 
direction in the (1,0,0)~Brillouin zone.
We remind here that the diffuse scattering has in the (1,0,0)-zone the butterfly shape \cite{xu} with the wings elongated
along $<1,1,0>$.

\begin{figure}[h]
\centering
\includegraphics[width=0.41\textwidth, angle=0]{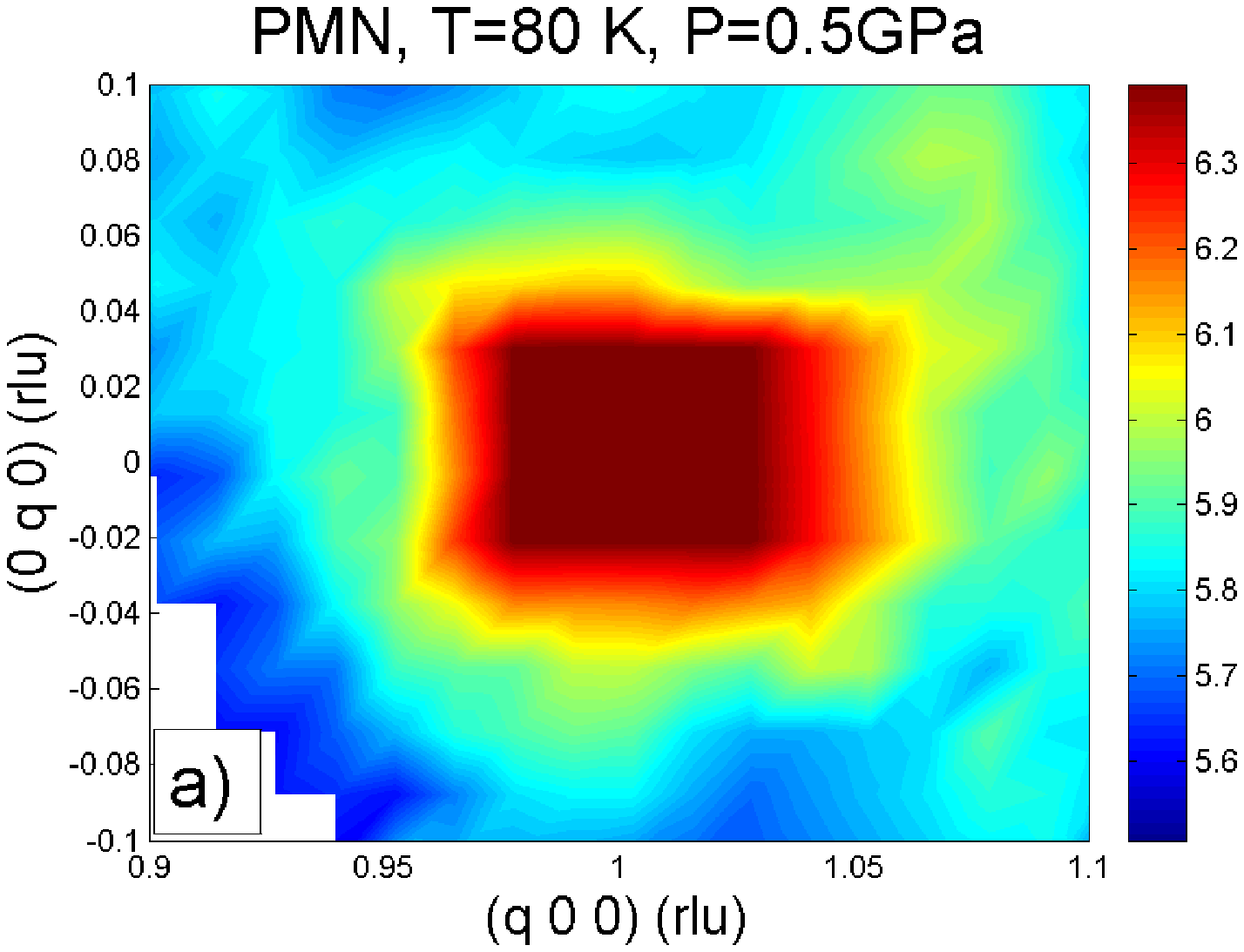}\hspace{2pc}%
\includegraphics[width=0.41\textwidth, angle=0]{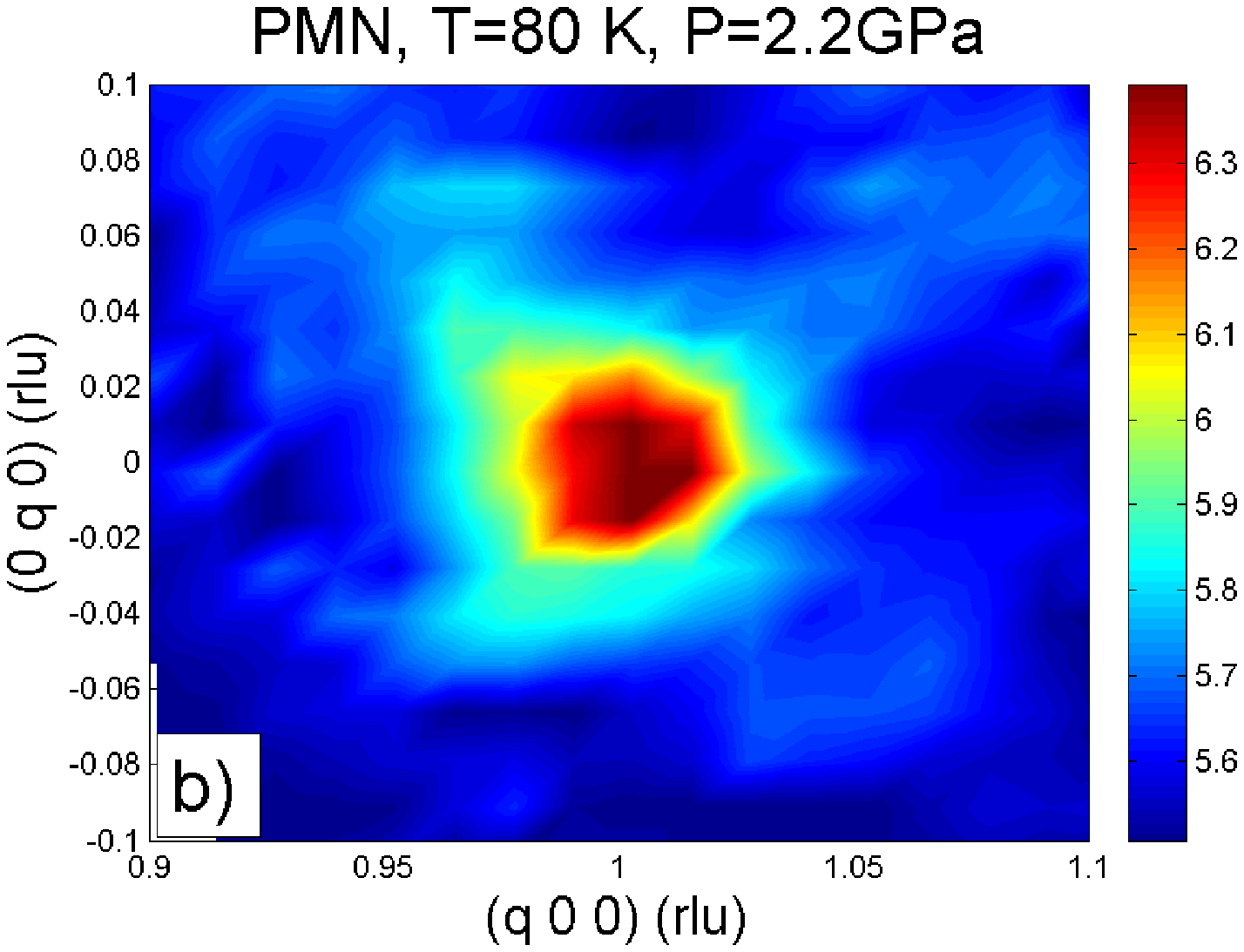}\hspace{2pc}%
\caption{Normalized diffuse scattering distribution in the (1,0,0) Brillouin zone measured at 80~K under 0.5~GPa-(a) and 2.2~GPa-(b).
 The diffuse scattering intensity is given in logarithmic scale.} 
\label{Fig1}
\end{figure}

\begin{figure}[h]

\centering

  \includegraphics[width=0.49\textwidth, angle=0]{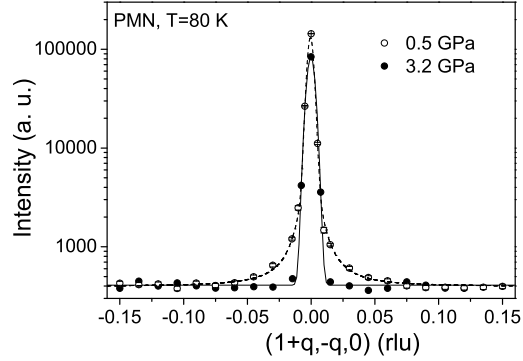}

  \caption{Constant Q-scans around the (1,0,0) Bragg peak in the [1,-1,0] direction measured at 0.5~GPa and 3.2~GPa respectively. 
The dashed and solid lines correspond to the fit (see text). Notice the pressure induced suppression of diffuse scattering.}
\label{Fig2}
\end{figure}

In Fig.~\ref{Fig1} the distribution of diffuse scattering around (1,0,0) is shown at 
0.5~GPa-(a) and 2.2~GPa (b). As one can see the butterfly shape is not observed. This is due to the high background
 resulting from the sample environment that makes the diffuse scattering indistinguishable at $|q|$ 
 bigger than 0.1~rlu. 
A better understanding  of the evolution of the diffuse scattering can be obtained by comparing the Q-scans at different pressures.
 In Fig.~\ref{Fig2}, for clarity only two scans at 0.5, and 3.2~GPa measured at 80~K are shown. The spectra contains the
Bragg peaks reflecting the long range order present at both pressures. The diffuse scattering decreases with pressure
 such that at  3.2~GPa only the resolution limited peak is left.
We have fitted the data assigning a Gaussian function for the Bragg peak and a Lorentzian function for the diffuse scattering.

\begin{figure}
\begin{center}
 \includegraphics[width=0.49\textwidth, angle=0]{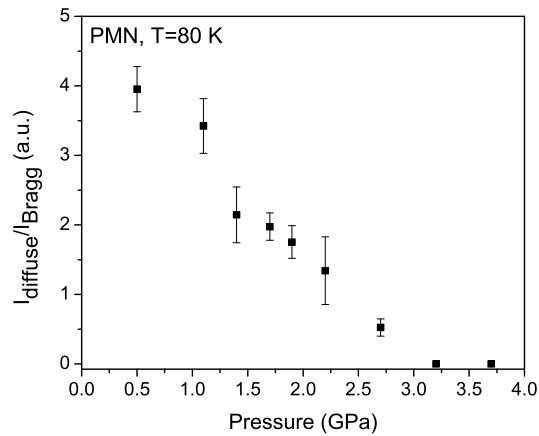}
\end{center}
\caption{\label{label}Diffuse scattering intensity measured along the [1,-1,0] direction in the (1,0,0) Brillouin zone, 
normalized to the Bragg intensity of the (1,0,0) reflection taken in two-axis mode  is plotted as a function of pressure. 
}
\label{Fig3}

\end{figure}

 A summary of the evolution of the diffuse scattering intensity as a function of pressure is shown in Fig.~\ref{Fig3}.
 Although we cannot resolve the pressure dependence of the correlation length of the short-range correlations, we do observe
 a decrease in the diffuse scattering. Hydrostatic pressure finally suppresses the polar short-range
 order around 3~GPa. 

In conclusion, our results indicate that there exists a strong correlation between atomic displacements, that give rise
to the PNR, diffuse scattering and the behavior of the dielectric permittivity in PMN. 
These results combined with the X-rays, neutron and dielectric studies provide further evidence that the local structure plays
an important role in the dielectric properties of PMN.

\ack{
This work was performed at the spallation neutron source SINQ, Paul Scherrer Institut, Villigen (Switzerland) and was
 partially supported by the Swiss National Foundation (Project No. 20002-111545).}
\section{References}

\smallskip

\end{document}